\documentstyle[12pt,aaspp4,amstex]{article}

\begin{document}

\title{Simulating Reionization in Numerical Cosmology}

\author{Aaron Sokasian}
\author{Tom Abel and 
Lars E. Hernquist} 
\affil{Department of Astronomy, Harvard University,
Cambridge, MA 02138\footnote{\tt asokasia@@cfa.harvard.edu,
tabel@@cfa.harvard.edu, lhernqui@@cfa.harvard.edu}}
\authoremail{asokasia@cfa.harvard.edu}

\begin{abstract}

The incorporation of radiative transfer effects into cosmological
hydrodynamical simulations is essential for understanding how the
intergalactic medium (IGM) makes the transition from a neutral medium
to one that is almost fully ionized.  Here, we present an approximate
numerical method designed to study in a statistical sense how a
cosmological density field is ionized by a set of discrete point
sources.  A diffuse background radiation field is also computed
self-consistently in our procedure. The method requires relatively few
time steps and can be employed with simulations having high resolution.  We
describe the details of the algorithm and provide a description of how
the method can be applied to the output from a pre-existing
cosmological simulation to study the systematic reionization of a
particular ionic species.  As a first application, we compute the
reionization of He {\small{II}} by quasars in the range $3\lesssim
z\lesssim6$.

\end{abstract}



\section{INTRODUCTION}

A reliable treatment of radiative processes in cosmological
hydrodynamical simulations is essential to understanding the reheating
and reionization of the intergalactic medium (IGM).  Besides enabling
a detailed modeling of the structure and evolution of the Ly $\alpha$
forest, which is now thought to originate from density fluctuations in
the low-density IGM (e.g. Cen et al. 1994; Zhang, Anninos \& Norman
1995; Hernquist et al. 1996), the coupling between ambient radiation
fields and baryons is essential for interpreting the nature of the
first generation of sources in the Universe.  Until recently,
radiative transfer effects have generally been treated by applying
local self-shielding corrections to an optically thin medium
(e.g. Katz et al. 1996; Gnedin \& Ostriker 1997) with only an
approximate treatment of the reprocessing of the radiation by the IGM
(e.g. Haardt \& Madau 1996).  Furthermore, at early times when the IGM
is still optically thick, radiation fields become increasingly
inhomogeneous and anisotropic (Reimers et al. 1997), and the inclusion
of radiative transfer effects becomes essential if one is to properly
track their evolution.  In fact, 3D radiative transfer calculations in
an expanding universe reveal that the propagation of ionization fronts
becomes quite complicated as the fronts expand from sources and
penetrate into voids before breaking through dense filamentary
structures (Abel, Norman \& Madau 1999; hereafter ANM).

In this paper, we describe an approach designed to model cosmological
reionization by combining 3D radiative transfer calculations with
outputs from hydrodynamical simulations.  Our
calculation of radiative transfer around point sources involves the
use of an algorithm based on a ray casting scheme similar to the one
described in ANM, implemented now in an evolving cosmological density
field.  Our method also includes a treatment of the ``diffuse''
component of the radiation field which is thought to become
significant once most of the volume is ionized.  As a first
application, we use our method to simulate how quasars reionize He
{\small{II}} in a cosmological density field.

\section{SIMULATING COSMOLOGICAL REIONIZATION}

In general, a description of the evolution of ionization zones around
cosmological sources requires a full solution to the radiative
transfer equation.  Such a solution yields everywhere the monochromatic
specific intensity of the radiation field in an expanding universe:
$I_{\nu}\equiv I(t,\vec{x},\hat{n},\nu)$, where $\hat{n}$ is a unit
vector along the direction of propagation of a ray with frequency
$\nu$.  Presently, it is computationally impractical to acquire a
complete, multi-dimensional solution for $I_{\nu}$ at the high
resolution required for cosmological simulations.  However, using a
sequence of well-motivated approximations, ANM have shown that it is
possible to reduce the dimensionality of the problem such that the
calculations become feasible.  Namely, the transfer equation is
reduced to a level where ionization rates can be computed on a
Cartesian grid using a ray casting scheme.  The ionization state of a
static density field can then be evolved through a series of time
slices, where rates are computed and neutral fractions are updated.
Although this method has been proven to be very successful for
computing the ionized regions surrounding multiple point sources, it
is still not a viable approach for calculating the systematic
reionization of an ionic species in a sizable volume containing a
large number of sources, which would require time-consuming
ray-casting calculations at each step.  Our approach relies on a new
algorithm which employs a simple jump condition to compute all
radiative ionizations from a given source in a single step.  This
removes the need to repeatedly re-cast rays and calculate rates at
every time step, thereby greatly speeding up the process.  Moreover,
our algorithm includes an approximate treatment of the diffuse
component of the radiation field.

As in ANM, we employ straightforward approximations to simplify the
calculations.  First, our radiative transfer calculations are done on
a uniform Cartesian grid whose scale $L$ will always be much smaller
than the horizon, $c/H(t)$, where $c$ is the speed of light and $H(t)$
is the time dependent Hubble constant.  This mitigates the need to
include Doppler shifting of frequencies in line transfer calculations.
Additionally, if the light crossing time $L/c$ is much shorter than
the ionization timescale, the time dependence of the intensities drop
out as well.  In the volumes we simulate this will certainly be true
and so we also make this approximation. (Note that this latter
assumption will inevitably break down near a source; however, this
problem can be remedied as shown below.)  Next, we assume that the
density field will experience negligible cosmological evolution during
the lifetime of single source.  This requires us to consider only
short-lived sources (at most a few $* 10^{7}$ years).  This assumption
allows us to perform all our radiative calculations during a source's
lifetime in a static density field, greatly reducing the complexity of
the algorithm.  Finally, in our most brazen assumption, we ignore
thermal feedback into the gas from radiative ionization, enabling us to
decouple our calculation of the radiation field from the
hydrodynamical evolution of the gas. This allows us to use
existing outputs from cosmological simulations to describe the
evolving density field during the reionization process. In reality, of
course, ionization introduces extra heat into the medium, and as
I-fronts move from small scales to large scales, there is a 
corresponding transfer
of power from small to large scales through nonlinear
evolution. This effect is somewhat accounted for in the underlying
cosmological simulation used in this paper which includes a uniform
ionizing background capable of heating the gas. However, one also
expects extra heating due to radiative transfer effects during the
reionization process, and such uniform backgrounds cannot reproduce
the observed increase in gas temperatures from the extra heating
(Abel \& Haehnelt 1999). As a result we expect systematic errors in
our solutions; however, the main purpose of this paper is to introduce
a method which can describe the general morphological evolution of the 
reionization process on large scales, which is fairly insensitive to such 
systematic errors. In particular, recombination rates depend weakly on
temperature and hence a proper accounting of photons is possible even
if the temperatures are not computed perfectly.

\section{IMPLEMENTATION}

The approach we take is to approximately describe the evolution of the
ionization state of the gas in a cosmological volume by iteratively
calculating the net effect of ionizing sources at regularly spaced
time intervals.  Density fields as well as information regarding
sources will be specified from outputs at desired redshifts from a
cosmological simulation.  The exact prescription for deriving this
information will be described in \S 4, where we discuss our first
application.

For simplicity, all sources are taken to have identical lifetimes.
Since we assume there will be a negligible amount of cosmological
evolution during a source lifetime, we compute the ionization state of
the gas in a static density field corresponding to the time at which a
source turns on.  To simplify bookkeeping, sources that would be active
during a particular time interval are assumed to all turn on at the
same redshift.  Thus, sources are binned together in time, with the
binsize being exactly one source lifetime.  Beginning with the
earliest redshift bin, the ionizing effects from member sources are
determined and ionization fractions are updated.  During the
calculations, the number of photons escaping into a diffuse background
component is tracked self-consistently.  The ionizing effect from this
background component is accounted for once all point source
calculations in the given time bin are completed.  If the source
lifetime is shorter than the time between successive outputs of the
density field, we use an interpolation scheme to update the density
field for the next bin of sources from the outputs.  In this way, the
evolution of the box is advanced by one source lifetime.  In \S 3.1
and \S 3.2 we describe the ray casting schemes used to carry out the
ionization calculations for point sources and the diffuse background
component respectively.  Then in \S 3.3, we describe how the
information derived from the ray casting schemes is used to set the
ionization fractions in affected cells.

\subsection{Point Sources}

In Figure 1, we give a block diagram for our algorithm, describing how
ionization from sources is calculated.  At a given redshift $z_i$, we
have a list of sources which are destined to switch on.  For each
source, we gather information regarding its location ($x,y,z$) and the
rate at which it emits ionizing photons for the species in question,
$\dot{N}_{ph}$ [$\text{s}^{-1}$].  A set of radial rays distributed
quasi-uniformly in solid angle is constructed for the source. These
rays will isotropically distribute ionizing photons from the source to
the intergalactic medium represented by the grid. A sufficient number
of rays is generated so that every cell at large distances from the
source is crossed by at least one ray. Each ray is ``cast'' into ray
segments according to how it intersects a cell boundary. The total
number of photoionizations is tracked along the ray and casting is
halted as soon as the total number of ionizing photons assigned to the
ray matches the cumulative number of absorptions (ionizations). During
the casting procedure, cells through which rays successfully traverse
are marked as {\it ionized} and their indices are stored. After all
rays from the source have been cast, ionization fractions are computed
for all cells marked {\it ionized} before casting starts for the next
source.

\subsubsection{Selecting Angles}

In this section we briefly describe the prescription for choosing the
angles that define the rays.  Our first step is to choose at least
$N_a=2\pi r/\Delta x$ rays so that at least one ray is cast into each
cell of side-length $\Delta x$ at the equator of a sphere with radius
$r$. To determine the required number of rays, we store the maximum
radius, $r_{max}$, needed to capture the furthest point of the
ionization front (I-front). We find that setting $r_{max}=\sqrt{3}\
L$, where $L$ is the length of the simulation box, gives us a safe
number of rays to properly describe the I-front within a periodic box
where rays can wrap around the box. Of course, in the completely
optically thin limit, ray path lengths may exceed $r_{max}$, but as we
discuss below where we describe our prescription for how rays
contribute to the background radiation field, we turn off periodic
boundary conditions for ray-casting long before the box has become
optically thin, thereby ensuring that path lengths remain less than
$r_{max}$.

To describe the rays we use spherical coordinates $(r,\phi,\theta)$,
\begin{equation}
\begin{align}
x&=r\cos \phi \cos \theta, \ \ \ 0\leq  \phi\leq 2 \pi \\
y&=r\sin \phi \cos \theta, \ -\frac{\pi}{2}\leq \theta \leq \frac{\pi}{2} \\
z&=r\sin \theta,
\end{align}
\end{equation}	     
where a discrete set of angles $(\theta_j,\phi_i)$ is used to define
the directional component of the rays which divide a sphere in
segments of similar areas. The discrete values of $\theta$ are given
by
\begin{equation}
\theta_j=(j-\frac{1}{2})\frac{2\pi}{N_a}-\frac{\pi}{2}, \ \ 1\leq j\leq\frac{N_a}{2}.
\end{equation}
Near the polar regions fewer azimuthal angles are required,
\begin{equation}
N^j_{\phi}=\text{max}(N_a\cos \theta,1),
\end{equation}
with the azimuthal angles chosen to be,
\begin{equation}
\phi_i=(i-\frac{1}{2})\frac{2\pi}{N_{\phi}}, \ \ 1\leq i \leq N_{\phi}.
\end{equation}
Each ray will thus describe an area (in units of $4\pi r^2$),
\begin{equation}
A_{\tiny{\theta_j,\phi_i}}=\frac{|\sin \theta_2-\sin \theta_1|}{2N_{\phi}^j},
\end{equation}
with the adjacent angles $\theta_1,\theta_2$ given by,
\begin{equation}
\begin{align}
\theta_1&=(j-\frac{1}{2})\frac{2\pi}{N_a}-\frac{\pi}{2}, \\
\theta_2&=j\frac{2\pi}{N_a}-\frac{\pi}{2}.
\end{align}
\end{equation}
Each ray is then assigned a photoionization rate proportional to the
area it describes on the sphere:
$A_{\tiny{\theta_j,\phi_i}}\dot{N}_{ph}$. This implementation allows
for a non-isotropic radiation field (such as bi-polar beaming) by, for
example, restricting the angles along which rays are cast. In such
cases, the photoionization rate assigned to each ray will be
$A_{\tiny{\theta_j,\phi_i}}\dot{N}_{ph} (4\pi/\Omega)$, where $\Omega$
is the total solid angle into which the radiation is beamed.

\subsubsection{Casting Rays}

Once angles have been chosen, rays are cast from the source into the
grid. The basic idea is to compute, for a given ray, the indices of
the grid cells it traverses, as well as the path lengths rays travel
within each cell. For a uniform, isotropic Cartesian grid such as the
one we will be using in our analysis, ray casting is straightforward
and we employ an implementation which is described in detail in \S 3.2
of ANM.

In general, casting for a particular ray is terminated once the total
number of ionizing photons assigned to it equals the total number of
photoionizations that have occurred along its path through the
grid. This process is described in detail in the following section.

\subsubsection{Ionizations}

For the purpose of studying the systematic reionization of a
particular ionic species, we shall primarily be concerned with the net
ionization effect of each source. In particular, we shall be
interested in calculating only the resultant I-front from all the
photons believed to have been released from the source during its
lifetime and we will be less concerned with the exact time dependent
nature of the expanding front. Furthermore, rather than describe the
ionization structure around a source when it has just turned off and
when there remains a substantial fraction of unabsorbed streaming
photons, we opt to allow each ray to first completely exhaust all its
ionizing photons before calculating the resulting structure.
Consequently, we will be unable to describe the I-front as it was {\it
exactly} at $z_i$ when the source turned off, but will rather focus on
how it appeared once all photons were absorbed. This allows us to
conserve all the photons emitted from the source. The elapsed time
between when the source turned off and when all photons are absorbed
will depend on the intensity of the source, its lifetime, and the
surrounding density field and corresponding ionization state.  In
general, we find that if the gas surrounding a source is optically
thick to the ionizing radiation, then the difference in time is short
compared to the time span over the reionization process occurs. In the
case of a source turning on in a mostly ionized region, the difference
can be a significant fraction of the total simulation time, but in
this case most rays reach the edges of the box without causing
dramatic changes to the already optically thin medium. Since we will
have a prescription for converting such unobstructed rays into a
background radiation field, this discrepancy will not significantly
affect our statistical study of the reionization process.

To calculate where along its path a ray exhausts its photon supply, we
simply integrate along its path length during the casting process
until the number of photoionizations of the species in question ($N$)
surpasses the number of ionizing photons ($N_{ph}$) carried by the ray
minus the total number of recombinations ($N_R$) occurring along its
path during the line transfer. More explicitly, for each ray
$(\theta,\phi)$, casting is terminated as soon as the following
inequality is satisfied,
\begin{equation}
N \geq \dot{N}_{ph} A_{\tiny{\theta,\phi}} t_{\tiny{life}} - N_R,
\end{equation}  
where $t_{\tiny{life}}$ is the source lifetime. This condition is
checked at the edge of each grid cell encountered by the ray.

The number of absorbing ions encountered by a {\it ray-cone} at a
distance $r$ from the source can be written as:
\begin{equation}
N=\int_0^r 4\pi  A_{\tiny{\theta,\phi}} r^2 n \ dr,
\end{equation}
where $n$ is the proper number density of the species in question at
redshift $z$ and $ 4\pi A_{\tiny{\theta,\phi}}r^2$ is the area
described by the ray at a distance $r$. We can easily calculate $N$
numerically at each intervening cell by summing over all the ray
segments $l$ in intervening cells,
\begin{equation}
N=\sum_{l}\  \frac{4\pi}{3} \ A_{\tiny{\theta,\phi}} \ n^l \ [S(l)^3-S(l-1)^3],
\end{equation}
where the radial length from the source to each cell crossing is given
by $S(l)$ (see ANM \S 3.2). Proper number densities are evaluated at each
intervening cell $i(l),j(l),k(l)$.

We now turn to the task of calculating $N_R$, the cumulative number of
recombinations along the path. For our analysis here, we shall assume
that each cell has a corresponding gas temperature, $T_i$, which will
be used to calculate recombination coefficients
($\alpha=\alpha(T)$). In calculating the cumulative number of
recombinations along the path, we need to take into account the fact
that there will be recombinations from the n=$\infty$ $\rightarrow$
n=$0$ transition which release extra ionizing photons that need to be
included in the radiative transfer calculations. Here we make the
assumption that the majority of these recombination photons
propagate unimpeded to the location of the I-front. This allows us to
use our existing radial rays to describe the propagation of these extra
photons. We therefore only use Case B recombination coefficients
(which exclude the n=$\infty$ $\rightarrow$ n=$0$ transition) in
calculating $N_R$. In light of equation (10), one can see that this
will allow the rays to penetrate further into the neutral medium due
to the extra emissivity implicitly included in $N_R^{Case B}$ as
opposed to $N_R^{Case A}$. Our approximate treatment for propagating
this extra diffuse component of the ionizing flux should not cause any
substantial departure from the true solution, especially since we are
dealing with R type I-fronts where the rate of recombinations is much
less than the rate of ionizing photons released from the source. It is
important to point out, however, that the calculations associated with
resultant ionization fractions (see \S 3.3) will involve the use of
Case A recombination coefficients since there we will require an
accounting of {\it all} recombination transitions.

The expression for $N_R$ in our ray casting scheme can then be
written as,
\begin{equation}
N_R=\sum_{l} \ \frac{4\pi}{3} \ A_{\tiny{\theta,\phi}} \ C^l_f \ n_e^l \ n_+^l\  \alpha_B \ [S(l)^3-S(l-1)^3]\ t(l,T_c),
\end{equation}
where $n_+^l$ and $n_e^l$ are the proper number densities of ionized
atoms and electrons respectively and $C^l_f$ represents the clumping
factor of the gas inside the cell. For the purposes of calculating
recombinations, we shall assume that all intervening cells have been
highly ionized ($\chi\approx 1$ where $\chi$ is the ionization
fraction) thereby fixing the values of $n_+^l$ and $n_e^l$.  The
function $t(l,T_c)$ represents the time elapsed in the cone segment
residing in cell $i(l),j(l),k(l)$ after it was ionized when a total
time of $T_c$ has passed since the source was turned on. We track
$T_c$ during the casting by summing over the crossing times through
each cell.  The crossing times, $t_s$, are calculated by integrating
the jump condition describing the propagation of the I-front radius
$r_{\tiny{I}}$ in a static density field (Abel 2000),
\begin{equation}
4\pi r^2_l n \frac{dr_{\tiny{I}}}{dt}=\dot{N}_{ph}-4\pi \alpha_B \int^{r_{\tiny{I}}}_0 n_e n_+ r^2\ dr.
\end{equation} 
The integral $\int^{t_s}_0 dt$ for a ray segment crossing a cell can
be expressed numerically as,
\begin{equation}
t_s^l=\frac{\frac{1}{3} n^l [S(l)^3-S(l-1)^3]}{\frac{\dot{N}_{ph}}{4\pi}-\sum_{l^{\prime}=0}^{l}\frac{1}{3} C_f^{l^{\prime}} n_e^{l^{\prime}} n_+^{l^{\prime}} \alpha_B [S(l^{\prime})^3-S(l^{\prime}-1)^3]},
\end{equation}
where the summation in the denominator accounts for the recombination
rate along the ray. In the case where $t_s^l$ is less than the light
crossing time across the pathlength through the cell,
$(S(l)-S(l-1))/c$, we simply reset $t_s^l$ to the light crossing
time. This allows us to avoid the unphysical effect of having faster
than light I-fronts near the source. In any case, arrival times to
each cell can be calculated by summing over individual crossing
times. The total elapsed time always includes a summation over all the
ray segments leading up to the edge of the ray (defined as ray segment
$l_c$) during the casting, such that,
\begin{equation}
T_c=\sum_{l=1}^{l_c} t_s^l.
\end{equation}
Figure 2 compares the results for arrival times computed using this
technique on a $100^3$ grid against what is expected analytically in
the case of a spherically expanding I-front (for hydrogen) in a
homogeneous density field. The radius and the spherical geometry are
perfectly recovered within the accuracy of the spatial resolution. The
specific parameters used in this test were, $\dot{N}_{ph}=10^{51}$,
$n_{\text{\tiny{H}}}=10^{-2}\ \text{cm}^{-3}$, cell-size $\Delta
x=0.15\ \text{kpc}$, and a constant Case B recombination rate for H
{\small{II}} $\rightarrow$ HI reactions of
$\alpha_{B_{\text{\tiny{H}}}}=3.6\times10^{-13}\ \text{cm}^3\
\text{s}^{-1}$ was selected. Finally we can write down our expression
for the elapsed time in each cone segment $t(l,T_c)$ in terms of the
total elapsed time $T$ at the edge of the ray at $l_c$,
\begin{equation}
t(l,T_c)=T_c-\sum_{l=1}^{l_c}t_s^l
\end{equation}
where $T_c$ and $l_c$ are updated at each new crossing.

It is important to point out that $T_c$ can exceed a source lifetime
before all photons have been absorbed from a given source. Regardless
of this fact, our interpolation scheme evolves the radiation field by
exactly one source lifetime at each step. In an effort to adhere to
this global time, we shall restrict the maximum time available for
recombinations in any cone segment to a single source lifetime.  In
particular, in the case where $T_c>t_{\tiny{life}}$, we count the
number of recombinations starting from the edge of the ray (ray
segment $l_c$) and work our way backwards along the ray until the
accumulated crossing times from the segments add up to a single source
lifetime. In this case, the summation in the denominator of (15) will
not start from $l^{\prime}=0$, but rather $l^{\prime}=l_{\star}$,
where $l_{\star}$ designates the ray segment marking the accumulation
of a source lifetime's worth of arrival times from segments spanning
backwards from the current location of the I-front. Similarly, the
summation in (13) will start from $l_{\star}$ resulting in an $N_R$
which excludes recombinations from segments very near the
source. Again, we do not anticipate that this will greatly alter the
global statistical properties of the reionization process. As stated
earlier, we find that during early times when most of the medium is
optically thick, the discrepancy between $T_c$ and $t_{\tiny{life}}$
is negligibly small. In the optically thin case, photons from the
source will inevitably reach the boundaries of the box and become part
of the diffuse background, where it is fair to approximate a single
source lifetime's worth of recombinations for such rays. Our treatment
will thus suffer when only a fraction of the medium in the box is
ionized, in which case $T_c$ is a few times larger than $t_s$ and only
some of the rays reach the boundaries. However, due to the rather
rapid onset of ionization overlap after just a few sources have been
active, the probability for many sources to turn on in a partially
ionized box is small.

\subsection{Diffuse Background Radiation Field}

As more sources turn on and progressively ionize the volume, there is
an increased likelihood that rays of ionizing radiation will loop
continuously through the box without encountering neutral pockets of
gas. In reality, this radiation becomes part of the diffuse
background. To simulate this background we adopt a simple prescription
for converting such perpetual rays into a uniform background which is
cast evenly from each side of the box. Our first step is to choose a
criterion to determine when source rays become part of the
background. Here we make the simplifying assumption that the intensity
of the background is correlated with the number of rays which reach
the boundaries of the box once the entire volume has surpassed a
certain level of ionization. In particular, once this global level of
ionization has been reached, we turn off the periodic boundary
conditions for the rays, strip the remaining photons from each exiting
ray reaching the boundary and add them to a photon reservoir
describing the background. The level of ionization when the background
receives these photons should be chosen such that it reflects a time
during the reionization process when a majority of the rays are able
to reach the boundaries of the box unobstructed. Although this may
seem somewhat arbitrary, the rapid onset of the overlap epoch ensures
that overall statistical properties of the reionization process remain
insensitive to this choice. In this paper, we find that a reasonable
choice is to turn on the background only after more than $50\%$ of the
volume has been ionized.

\subsubsection{Casting Rays}

Typically, cosmological simulations treat diffuse radiation fields as
being uniform so that the intensity of the radiation is constant
everywhere. Although this is a good approximation in the limit when
the entire box is optically thin, it can fail to reproduce the natural
way in which pockets of neutral gas are ionized from the outside
in. Since in our simulations the background can gain in intensity
while the volume is still partially neutral, we adopt a geometric
scheme for casting background rays which can describe the ionization
of pockets of neutral gas while retaining an isotropic nature.  The
scheme we choose has the geometric property of casting parallel rays
inwards from each face of the box. In particular, one ray from each
cell face is cast perpendicularly from its corresponding box
face. Thus for a box with a resolution of $N_c$ cells in each
dimension, the number of background rays generated at each step is
$6\times N_c^2$.  In essence, the background can be thought of as being
comprised of six distinct {\it planar} sources.

\subsubsection{Ionizations}

After the criterion for allowing rays to contribute to the background
has been satisfied, the number of {\it escaping} photons is recorded
during each redshift bin when sources are turning on. This is then
added to a global variable, $N_{ph}^{bk}$, representing the total
number of photons available to the background. $N_{ph}^{bk}$ is
allowed to change as the background exhausts its photon supply or
gains more photons from sources at each step. Additionally, to account
for the redshifting of background photons below the ionization
threshold frequency $\nu_I$, at each redshift step $z_i$ the new
number of ionizing photons present in the diffuse background is
adjusted according to:
\begin{equation}
{N_{ph}^{bk}}^{\prime}=N_{ph}^{bk}\frac{\int_{\nu_I^{\prime}}^{\infty}f(\nu)\nu^{-1}d\nu}{\int_{\nu_I}^{\infty}f(\nu)\nu^{-1}d\nu},
\end{equation}
where $f(\nu)$ is the spectral profile of the specific intensity of
the background with
$\nu_I^{\prime}=\nu_I(1+z_i)/(1+z_{i-1})$. Assuming the sources
responsible for the diffuse background are all described by the same
power-law spectral index $\alpha$ beyond $\nu_I$ and that there is
negligible reprocessing of this radiation by the IGM, equation (18)
reduces to
\begin{equation}
{N_{ph}^{bk}}^{\prime}=N_{ph}^{bk}\biggl(\frac{1+z_{i-1}}{1+z_i}\biggl)^{-\alpha}.
\end{equation}

Ionizations from the background are computed by assigning to each
background ray ${N_{ph}^{bk}}^{\prime}/(6\times N_c^2)$ ionizing
photons and tracking how far it can penetrate into the volume. This is
accomplished by following the same prescription for tracking source
rays only now the geometry is much simpler. In particular, the number
of absorbing atoms one background ray will intercept during its course
through $l$ cells will simply be,
\begin{equation}
N=\sum_{l}\ n^l V_{cell},
\end{equation}
where $V_{cell}$ is the cell volume. Similarly, the number of
recombinations along the same path will be given by,
\begin{equation}
N_R=\sum_{l} \ C^l_f n_e^l n_+^l \alpha_B V_{cell}\ t(l,T_c),
\end{equation}
where $t(l,T_c)$ now represents the elapsed time spent in cell
$i(l),j(l),k(l)$ when a total time of $T_c$ has passed since the ray
was cast from the edge of the box. The crossing times implicit in
$T_c$ for background rays are given by,
\begin{equation}
t_s^l=\frac{N}{\frac{{N_{ph}^{bk}}^{\prime}}{6N_c^2t_s}-\sum_{l^{\prime}=0}^{l}
C_f^{l^{\prime}} n_e^{l^{\prime}} n_+^{l^{\prime}} \alpha_B V_{cell}},
\end{equation}
where the term ${N_{ph}^{bk}}^{\prime}/6N_c^2t_s$ represents the rate
of ionizing photons streaming along the background ray at each time
step ($t_s$).

\subsection{Computing Ionization Fractions}    

Within each redshift bin, ionization fractions for the entire grid are
updated after each source has been handled, with point sources always
preceding plane sources (the latter being the six faces comprising the
background). To accomplish this in a self-consistent manner, we will
need to introduce a set of variables for each cell. These variables
will store various parameters needed to calculate the cumulative
effects from all sources in a given redshift bin.

We start by describing the role of the boolean variable $Q$, which is
used to distinguish whether a particular cell has been ionized by a
given source.  At the start of the ray casting calculation for a
source, all cells in the grid are initialized with $Q=false$. During
the casting, if a ray successfully traverses a given cell without
exhausting its photon supply (eq. 10), then it is flagged as an {\it
ionized} cell via the assignment $Q=true$. If the ray responsible for
flagging the cell originates from a point source then the proper
distance from the source to the cell is recorded in the variable
$r_s$, which is used to calculate the photoionization rate. After ray
casting calculations are completed for each source, only marked cells
will have their ionization fractions updated.

In a given redshift bin where multiple sources are due to switch on,
we shall need to guard against the possibility of overestimating the
number of recombinations in a cell which is affected by multiple
sources.  That is, we will restrict the number of recombinations
allowed in each cell through which rays from different sources
pass. We achieve this by allowing recombinations to occur only along
rays that pass through cells which have not been affected by sources
other than the one in question. We thus introduce for each cell a new
integer variable $R$, which will store the ID number of that source
whose rays first penetrate it. For a ray to suffer recombinations
within a given cell therefore requires that $R$ be equal to either the
source ID, or to zero (the cell is virgin).  The recorded values are
kept in memory until the last source in a given redshift bin has
finished its ray casting calculations. In each new redshift bin,
values of $R$ are reset to zero everywhere. Though our approach for
restricting recombinations will be dependent on the order in which
point sources in a given bin turn on, over the course of the
simulation we expect to statistically reproduce the correct
behavior. In the case of plane sources, the order in which each face
is turned on is randomized at each redshift bin to avoid producing
preferential ionizations along certain sides of the box during the
course of the calculation.

After ray casting is completed for a given source, new ionization
fractions are set for all cells marked $Q=true$ according to the
equilibrium condition,
\begin{equation}
C_f n_e^{\prime} n_+^{\prime} \alpha_A=\Gamma_{ph}n^{\prime} +
C_f \Gamma_{coll}n^{\prime}n_e^{\prime},
\end{equation}
where $\Gamma_{ph}$ and $\Gamma_{coll}$ are the photoionization rate
and the temperature dependent collisional ionization rate for the
ionic species in question, respectively. Note that we have used a Case
A recombination coefficient in our equilibrium condition since we are
interested in accounting for all recombination transitions for regions
that have been ionized. We shall discuss the details associated with
extracting the temperature dependent terms $\alpha_A$, $\alpha_B$ and
$\Gamma_{coll}$ as well the clumping factor $C_f$ from the
cosmological simulation below.  Note that the equilibrium condition
above is not necessarily achieved in highly underdense regions where
chemical timescales exceed the Hubble time; however, this will have a
negligible effect on the overall morphological description of the
ionization process, which is the primary objective of our type of
study. In any case, the above equation can then be solved for the
ionization fraction $\chi$ of the species.

The photoionization rate in a cell due to radiation from a point
source is given by,
\begin{equation}
\Gamma_{ph}=\frac{\bar{\sigma}\dot{N}_{ph}}{4\pi r_s^2} \
\text{s}^{-1},
\end{equation}
where $\bar{\sigma}$ is the mean cross section for photoionization. In
the case of a plane source, the photoionization rate is expressed as,
\begin{equation}
\Gamma_{ph}=\frac{N_{ph}^{bk} \ \bar{\sigma}}{6 N_c^2 A_c t_s} \
\text{s}^{-1},
\end{equation}
where $A_c$ is the proper area of a cell face. Our expression for the
mean cross section for photoionization is computed for a known
radiation spectrum of the sources $J(\nu)\propto\nu^{-\alpha}$ according to,
\begin{equation}
\bar{\sigma}= \frac{\int_{\nu_I}^{\infty} \frac{J(\nu)}{h\nu} \
\sigma(\nu)\ d\nu}{\int_{\nu_I}^{\infty} \frac{J(\nu)}{h\nu}\ d\nu} \
\text{cm}^2,
\end{equation}
where $\sigma(\nu)$ is the value of the cross section at frequency
$\nu$.

Since in a single redshift bin, a cell may be affected by multiple
sources, the photoionization rate $\Gamma_{ph}$ will have to be
updated for each source. At each redshift bin, we initialize
$\Gamma_{ph}=0$ everywhere on the grid. Once ray casting is completed
for a given source (point {\it or} planar), the photoionization rate
in each marked cell is updated according to,
\begin{equation}
\Gamma_{ph}^{new}=\Gamma_{ph}^{old}+\Gamma_{ph}^{add},
\end{equation}
where $\Gamma_{ph}^{add}$ is the additional contribution to the rate
from the current source.

\section{AN APPLICATION}

We now describe how the above method can be used to analyze outputs
from a cosmological simulation to study the systematic reionization of
He {\small{II}} from quasars. We will use information from coarsely
spaced outputs of a cosmological simulation to interpolate density
fields for the ionization calculation. Source information will also be
derived from the cosmological outputs with the aid of a particle group
finder. Figure 3 gives the block diagram which schematically describes
how the analysis proceeds.

The cosmological simulation we will use in our analysis is based on a
smoothed particle hydrodynamics (SPH) treatment, computed using the
parallel tree-code GADGET developed by Springel, Yoshida \& White
(2000).  The particular cosmology we examine is a $\Lambda$CDM model
with $\Omega_{b}=0.04$, $\Omega_{DM}=0.26$, $\Omega_{\Lambda}=0.70$,
and $h=0.67$ (see, e.g., Springel, White \& Hernquist 2000).  The
simulation uses $224^{3}$ SPH particles and $224^{3}$ dark matter
particles in a $67.0 \ \text{Mpc}/h$ comoving box, resulting in mass
resolutions of $2.970\times 10^8 \ M_{\odot}/h$ and $1.970\times 10^9
\ M_{\odot}/h$, respectively.

\subsection{Calculating Density Grids}

Since our ray casting scheme is designed within the context of a
uniform Cartesian grid, the gas represented by SPH particles will
first have to be distributed into cells to form the required density
fields. Density grids will have to be produced for each available
cosmological output in the redshift range of interest, from which we
will interpolate the dynamical evolution of the gas for intermediate
steps for our ionization calculation. In our analysis, we will be
interested in the redshift range $3\lesssim z\lesssim6$ where He
{\small{II}} reionization is believed to have occurred. The
cosmological simulation we use provides us with roughly $20$ data
snapshots in this range.

We would like to retain as much of the SPH resolution as possible and
therefore employ grids which have a total number of cells comparable
to the number of gas particles used in the cosmological simulation. In
this paper, we choose a conservative resolution of $200^3$ for our
grids. The basic idea behind gridding is to properly distribute the
mass from the particles into cells according to how the volume
associated with each particle overlaps the cells in its
vicinity. Since the effective size of an SPH particle can vary
according to its surrounding density, particles will distribute their
mass over a varying number of cells.

Our prescription for distributing the mass relies on using the
smoothing kernel associated with the SPH particles.  In particular,
the density contribution from a particle into an overlapping cell is
obtained by averaging over the density kernel of the particle at
discrete points uniformly distributed within the cell in question. The
number of points over which the kernel density is averaged is directly
related to the accuracy of the mass distribution. However, due to the
very large number of particles through which one must loop during the
gridding process, it becomes computationally expensive to use many
points.  We found that just $8$ points within each cell gives a
surprisingly good description of the distribution.  To conserve mass
in light of the limited accuracy of this scheme, we normalize the
total mass allotted in overlapping cells to equal the total mass of
the contributing particle. In the case where a particle is entirely
contained within one cell, the particle contributes all its mass to
that cell, in the form of a uniform density over the cell's volume.

In an effort to retain information associated with the mass
distribution on scales smaller than a cell, we calculate an overall
clumping factor for each cell based on the clumping characteristics of
the particles contained within. In particular, we calculate the volume
averaged clumping factor in each cell according to,
\begin{equation}
C_f=\frac{\langle \rho^2 \rangle}{\langle \rho
\rangle^2}=\frac{\biggl(\sum_i\ \rho_p(i)^2 \
\frac{m_p(i)}{\rho_p(i)}\biggl)}{\biggl(\sum_i \ \rho_p(i) \
\frac{m_p(i)}{\rho_p(i)}\biggl)^2}\sum_i \ \ \frac{m_p(i)}{\rho_p(i)},
\end{equation}
where the summations are over particle indices whose center positions
are located within the cell. The variables $\rho_p(i)$ and $m_p(i)$
represent the quoted density and mass of particle $i$ respectively,
with $\rho_p(i)/m_p(i)$ giving the corresponding particle volume. It
is obvious that the clumping factor for a given cell can only be
greater than unity if it contains more than one particle. In the case
where a cell does not contain any particles, its clumping factor is
set to unity. It is important to point out that the above estimate
assumes that all high density gas is uniformly ionized. As a result
we expect that we will be overestimating the clumping factor
significantly in cells which harbor highly dense compact systems.
This represents one of the drawbacks of using a uniform Cartesian grid
to describe our density field, nevertheless, due to the comparable
level of resolution to the underlying cosmological simulation, we
expect only a small fraction of the total cells in our grid to
significantly overestimate the clumping factor. Furthermore, such
cells generally harbor the sources themselves in which case we simply
set the ionization fraction to unity within the virial radius of the
mass group. For the grid resolution chosen in this paper, the virial
radius for a typical source is about the width of a cell.

In a similar fashion, recombination rates and collisional ionization
rates for a given cell are calculated on a particle by particle basis.
In particular, rates are calculated by averaging over particle
contributions according to their temperatures.  It is is important to
restate at this point that the underlying cosmological simulation we
use in this paper neglects to include radiative transfer effects in
its calculations of the photoheating rate, thereby substantially
underestimating the energy input during and after the reionization
(see Abel \& Haehnelt 1999). As a result, we expect that the
corresponding gas temperatures of the particles to be significantly
lower than expected. To compensate for this shortcoming we artificially
set the temperature of all the gas which becomes fully ionized in
helium to $T=2.0\times10^4\ K$ unless its temperature was greater
beforehand, in which case the prior temperature is used to calculate
rates. In this paper the fitting formulae for the temperature
dependent He {\small II} recombination rates (cases A and B) is taken
from Hui \& Gnedin (1997) (based on a fit to the data from Ferland et
al. (1992)) and the fitting formula for the collisional ionization
rate of He {\small II} is taken from Black (1981).

Once all the particles from a cosmological output have been processed,
the resulting grid including clumping factors, and rates is written to
a file and stored.  These files are used to interpolate the dynamic
evolution of the gas during the ionization simulation.

\subsection{Source Selection}

The Gunn-Peterson constraint (Gunn \& Peterson 1965) on the amount of
neutral material along the line of sight to distant objects requires
that the hydrogen component of the intergalactic gas to have
become highly
ionized by $z\approx 6$ (Fan et al. 2000) and that the helium
component was ionized by $ z\approx 2.5$ (Davidsen, Kriss, \& Zheng
1996). There has been much debate over the nature of the sources
responsible for ionizing the gas; namely whether they were stellar in
nature or quasars. However, a discerning feature of He {\small{II}} is
its relatively high ionization potential, requiring that the sources
responsible for ionizing this component of the intergalactic gas had
intrinsically hard spectra. This supports the idea that He
{\small{II}} was primarily ionized by quasars which have much harder
spectra than stellar sources. In particular, the ratio of He
{\small{II}} to H {\small{I}} Lyman continuum photons from
star-forming galaxies is only about 2\% as compared to 10\% for a
typical quasar (Leitherer \& Heckman 1995). Moreover, observations of
helium absorption (Jakobsen et al. 1994; Davidsen, Kriss, \& Zheng
1996; Hogan, Anderson, \& Rugers 1997; Reimers et al. 1997; Heap et
al. 2000; Smette et al. 2000) suggest that the optical depth does not
steadily increase with lookback time, but instead drops sharply around
$z\approx 3$ when quasars were prevalent.  In our analysis we will
consider quasars as the sole sources of He {\small{II}} ionizing
radiation.

In this section we describe how candidate quasars are selected from
the cosmological simulation to act as sources of ionizing
radiation. This will involve identifying dense clumps of gas particles
that stand out from the background as galaxies which are plausible
quasar hosts, and adopting a prescription for selecting
a subset of these objects as actual sources according to an empirical
quasar luminosity function.

Although it will be possible to interpolate gridded density fields
between these redshifts, source information will be available only at
the cosmological outputs where one can use an algorithm designed to
employ particle information to identify collapsed, highly overdense
objects. As a result, during the course of the ionization evolution,
source information will be based on the nearest former cosmological
output. Since highly collapsed objects harbor relatively large amounts
of matter, their large inertia causes them to remain relatively
stationary over the course of a single light crossing time, making our
approximate treatment for identifying their locations during the
intermediate steps very reasonable.

To compile the list of sources expected to turn on during the
simulation we follow the algorithm represented by the block diagram in
Figure 4. The first task required is to store in descending order an
array of redshifts, $z_{i}$, corresponding to intervals of one source
lifetime. These redshifts will designate when sources turn on and off
during the ionization calculation. Next, we loop over the cosmological
simulation outputs, hereafter referred to as hydro outputs, to
identify plausible sources. Due to the high resolution of the
simulations we are restricted to employing a fast algorithm capable of
searching through a large dataset for dense clumps containing many
particles. For our analysis we employ a Friends-of-Friends group
finder which is designed to efficiently handle groups with very large
numbers of particles ($>10^{6}$), where it becomes imperative that the
number of pair comparisons is kept to an absolute minimum. To identify
galaxy type objects, we specify a maximum linking length corresponding
to a group overdensity of roughly $200$. Additionally, we specify a
minimum mass, $M_{min}$, required for a group to be considered a
source. For every possible source identified, the center of mass
location and total gas mass of the group are recorded and stored in
arrays.

Once a list of plausible sources for a given hydro output has been
compiled, all identified groups are logarithmically binned by mass and
a tally of the number of groups in each bin is made. The results can
then be fitted to form an analytic representation of the mass
function. Here we use a simple power-law of the form:
\begin{equation}
\frac{dN}{d\log{M}}=10^{(a\log{M}+b)}
\end{equation}   
where $dN$ is the number of groups with total gas masses between
$\log{M}$ and $\log{(M+dM)}$, and $a$,$b$ are fitted parameters. Our
analytic representation is motivated by the general power-law nature
of theoretical halo mass functions at high redshift (see Press \&
Schechter 1974). It is important to point out that the above mass
function applies only to the range of group masses found in our
simulation volume. It is thus imperative that our cosmological
simulation be large enough in scale and also have sufficient
resolution so that we sample a representative range of group (source)
masses. As we shall discuss later in this section, the cosmological
simulation used in this paper meets these criteria by being able to
reproduce the observed range of quasar luminosities with a constant
mass-to-light ratio.

Our next step is to define a selection criterion which will determine
a realistic subset of sources that will actually be activated during
the calculation of the radiation field.  In particular, we would like
to select sources according to the observed quasar luminosity function
$\phi(L,z)$ (LF). Here, $\phi(L,z)dL$ is the number of quasars
per unit comoving volume at redshift $z$, with intrinsic luminosities
between $L$ and $L+dL$.  Given the observational data, the inferred
luminosity function will depend upon the assumed cosmology and quasar
spectra. In our simulations, we will adopt the double-power-law model
presented by Boyle et al. (1988) using the open-universe fitting
formulae from Pei (1995) for the $B$-band (4400 \AA \ rest-wavelength)
LF of observed quasars, with a rescaling of luminosities and volume
elements for our $\Lambda$CDM cosmology. The parametric form of the
double power-law LF can be written as
\begin{equation}
\phi(L,z)=\frac{\Phi_{\star}/L_{z}}{(L/L_{z})^{\beta_1}+(L/L_z)^{\beta_2}},
\end{equation}
where the break luminosity, $L_{z}$, evolves with redshift
according to:
\begin{equation}
L_{z}=L_{\star}(1+z)^{\alpha-1}\mbox{exp[$-(z-z_{\star})^2/2\sigma_{\star}^2$]}.
\end{equation} 

In this model, the evolution of quasars peaks at $z_{\star}$ with a
characteristic dispersion of $\sigma_{\star}$. The redshift factor
$(1+z)^{\alpha-1}$ represents the explicit dependence of the spectral
index $\alpha$ where a UV spectrum, $f(\nu)\propto\nu^{-\alpha}$, has
been assumed.  The values of the fitting parameters are
$z_{\star}=2.77$, $\sigma_{\star}=0.91$, $\beta_1=1.83$,
$\beta_2=3.70$, $log (\phi_{\star}/\mbox{Gpc}^{-3})=2.37$, and $log
(L_{\star}/L_{\odot})=13.42$ for an open-universe cosmology with
$h=0.50$, $q_o=0.1$, and $\alpha=1.0$. We use the above values and
convert the LF to our $\Lambda$CDM cosmology by rescaling the
luminosity and volume element for each redshift of interest. Our
conversion of the LF ignores any spectrum-dependent $k$-corrections
which are thought to be small assuming an $\alpha=1.0$ power law for
the quasars (Haiman \& Loeb 1998). It is important to point out that
this model predicts that a large fraction of the luminosity at $z>2$
arises from quasars that have not yet been observed.  With a fit to
the LF which goes as $\phi(L)\propto L^{-1.83}$ at the faint end, the
emissivity, $\int \phi(L,z)\ LdL$ (in ergs $\mbox{s}^{-1}$
$\mbox{Hz}^{-1}$ $\mbox{cm}^{-3}$), converges only as $L^{0.17}$, and
it becomes apparent that a large portion of the total emissivity will
arise from this regime.  Nevertheless, the extrapolation of the LF to
fainter quasars seems to be reasonable based on the analysis of Haardt
\& Madau (1996) where they use the above LF and a realistic form for
quasar spectra to calculate the intensity of the ionizing background,
$J$. Using the cosmological radiative transfer equation for diffuse
radiation (e.g. Peebles 1993),
\begin{equation}
\biggl(\frac{\partial}{\partial
t}-\nu\frac{\dot{a}}{a}\frac{\partial}{\partial
\nu}\biggl)J=-3\frac{\dot{a}}{a}J-c\kappa J+\frac{c}{4\pi}\epsilon,
\end{equation}
where $a$ is the scale factor, $\kappa$ is the continuum absorption
coefficient per unit length along the line of sight, and $\epsilon$ is
the proper space-averaged volume emissivity, Haardt \& Madau include
the effects of absorption and emission by intervening clouds to
calculate the evolution of $J$. They show that $J_{912\ \text{\AA}}$
increases from $\approx 10^{-23}$ ergs $\mbox{s}^{-1}$
$\mbox{cm}^{-2}$ $\mbox{sr}^{-1}$ $\mbox{Hz}^{-1}$ at the present
epoch to $\approx 5\times 10^{-22}$ ergs $\mbox{s}^{-1}$
$\mbox{cm}^{-2}$ $\mbox{sr}^{-1}$ $\mbox{Hz}^{-1}$ at $z=2.5$. This
result is very consistent with high-resolution studies of the
proximity effect which give $J_{912\ \text{\AA}}\approx 5 \times
10^{-22}$ obtained for a redshift range $z=1.7-4.1$ (Giallongo et
al. 1996) and $J_{912\ \text{\AA}} \approx 10 \times 10^{-22}$ (Cooke
et al. 1997) for a similar redshift range.  It is furthermore
consistent with limits imposed by the opacity of the Lyman alpha
forest at both high (e.g. Rauch et al. 1997) and low (e.g.  Dav\'e et
al. 1999) redshifts.

To be able to utilize the LF in our selection process, we will first
need to prescribe some way of assigning luminosities to our list of
plausible sources. In this paper we will simply assume that all
sources have $B$-band luminosities directly proportional to their gas
mass. This introduces a constant mass-to-light ratio,
\begin{equation}
\frac{M}{L_{B}}=\xi,
\end{equation} 
as another parameter in the analysis.

We begin the selection process by first determining the number of
sources expected at redshift $z_{i}$ with luminosities above
$L_{min}$,
\begin{equation}
N_{e}(z_{i},L_{B}>L_{min})=\int_{L_{min}}^{\infty}\phi(L_{B},z_{i})\
dL_{B}.
\end{equation}  
Here, the minimum luminosity, $L_{min}$, is defined as the $B$-band
luminosity corresponding to $\xi/M_{min}$. $N_{e}$ will thus designate
the total number of sources chosen at redshift $z_{i}$ during the
selection process. To convert $N_{e}$ to an integer in an impartial
manner, a random number between $0$ and $1$ is generated and compared
to the fractional component of $N_{e}$. If the fractional component is
larger than the generated number, $N_e$ is rounded up to the nearest
integer, else $N_{e}$ is rounded down.

Next, we loop over every plausible source to determine whether it will
be chosen. Our selection criterion will consist of assigning a random
number between $0$ and $1$ to the candidate source of mass $M$ and
selecting it if the assigned value matches or falls below the value of
a some probability function $P(M)$. To ensure that we match the quasar
luminosity function, we define the probability function to be:
\begin{equation}
P(M)=\biggl(\frac{dN}{dM}\biggl)_{LF}\biggl(\frac{dN}{dM}\biggl)_{Hydro}^{-1},
\end{equation} 
where $(dN/dM)_{LF}$ is the expected source mass function for a
simulation box with comoving volume, $V_{box}$, as derived from the
luminosity function:
\begin{equation}
\biggl(\frac{dN}{dM}\biggl)_{LF}=\xi^{-1}\phi(M/\xi,z_{i})V_{box},
\end{equation} 
and $(dN/dM)_{Hydro}$ is just our analytic representation of the mass
function attained from the actual hydro simulation now rewritten as:
\begin{equation}
\biggl(\frac{dN}{dM}\biggl)_{Hydro}=M^{-1}10^{(a\log{M}+b)}.
\end{equation} 
The loop is terminated when exactly $N_{e}$ sources have been
selected. If the number selected falls short of $N_{e}$ after one run
through the list of plausible sources, the list is reordered randomly
and source selection is continued from the beginning of the list.  In
this manner, a list of sources is compiled for every redshift step of
the simulation, $z_{i}$, with plausible source lists and mass
functions being updated near every hydro output redshift.

Once a source with mass, $M$, has been selected, it is assigned a
$B$-band luminosity, $L_{B}=M/\xi$ (in ergs $\mbox{s}^{-1}$). This
luminosity along with an assumed spectral form, is then used to
compute the amount of He {\small{II}} ionizing flux that will be
generated while the source is active.  In this paper, we shall assume
for all sources, a multi-power-law form for the spectral energy
distribution, adopted from Madau, Haardt, \& Rees (1999) for quasars,
\begin{equation}
L(\nu)\propto
\begin{cases}
\nu^{-0.3} & (2500\ \mbox{\AA}<\lambda<4400\ \mbox{\AA});\\ \nu^{-0.8}
& (1050\ \mbox{\AA}<\lambda<2500\ \mbox{\AA});\\ \nu^{-1.8} &
(\lambda<1050\ \mbox{\AA}),
\end{cases}
\end{equation}
where the different slopes have been continuously matched. This is
based on the rest-frame optical and UV spectra of Francis et
al. (1991), Sargent, Steidel, \& Boksenberg (1989), and the EUV
spectra of radio-quiet quasars (Zheng et al. 1998), and is very
similar to the spectrum assumed in Haardt \& Madau (1996) except for
the slightly steeper index at shorter wavelengths. It is important to
bear in mind that there is significant dispersion in each of the
spectral indices used to construct the composite spectrum which could
influence our final results. To calculate the amount of He
{\small{II}} ionizing flux, we will need an expression for the
intensity at wavelengths short of the ionization threshold ($228$
\AA). Using the above spectral form, we thus derive the following
expression for the specific intensity in the regime $\lambda<1050$
\AA:
\begin{equation}
\begin{align}
J_s(\nu)&=J_s(\nu_{\text{\tiny{$B$}}})\biggl(\frac{4400}{2500}\biggl)^{-0.3}\biggl(\frac{2500}{1050}\biggl)^{-0.8}\biggl(\frac{\nu}{\nu_{\text{\tiny{$1050$}}}}\biggl)^{-1.8}\\
&\approx 0.423\
J_s(\nu_{\text{\tiny{$B$}}})\biggl(\frac{\nu}{\nu_{\text{\tiny{$1050$}}}}\biggl)^{-1.8},
\end{align}
\end{equation}
where $\nu_{\text{\tiny{$B$}}}$ and $\nu_{\text{\tiny{$1050$}}}$ are
frequencies evaluated at $4400$ \AA\ and $1050$ \AA\ respectively, and
$J_s(\nu_{B})$ is the source intensity evaluated at frequency
$\nu_{B}$ (in ergs $\mbox{s}^{-1}\ \mbox{cm}^{-2}\ \mbox{sr}^{-1}\
\mbox{Hz}^{-1}$). Next, we calculate $J_s(\nu_{B})$ using the assigned
$L_{B}$ for the source,
\begin{equation}
J_s(\nu_{B})=\frac{L_{B}}{4\pi A_{s} \nu_{B}},
\end{equation}
where we have used the $\nu L_{\nu}$ definition for band luminosities,
and we have assumed an isotropically emitting surface with area $A_s$
for our source (though this will inevitably drop out of our
calculations).  We then use our expression for $J_s(\nu)$ to calculate
the number of ionizing He {\small{II}} photons released per second by
the source,
\begin{equation}
\begin{align}
\dot{N}_{ph} & =\int_{\nu_I}^{\infty}\frac{4\pi A_{s}J_s(\nu)}{h\nu}\
d\nu\\ &\approx0.423 \ \frac{L_{B}}{h\nu_B}
\int_{\nu_I}^{\infty}\frac{1}{\nu}\biggl(\frac{\nu}{\nu_{1050}}\biggl)^{-1.8}d\nu\\
&\approx 2.44\times10^{52} \ \biggl(\frac{L_{B}}{L_{B,\odot}}\biggl)\
\mbox{s}^{-1},
\end{align}
\end{equation}
where $h$ is Planck's constant, $\nu_I$ is the ionization threshold
frequency for He {\small{II}}, and $L_{B,\odot}$ is the solar $B$-band
luminosity (in ergs $\mbox{s}^{-1}$). An ionization rate is computed
and recorded for each source immediately after it is selected.

At the end of the entire source selection process, we will have
recorded a list of source locations and ionization rates for every
redshift step of the simulation. In selecting the sources and
computing their intensities we have introduced three parameters: (1) a
universal source lifetime, (2) a minimum source mass, and (3) a source
mass-to-light ratio.  For our first application, we have adopted a
fiducial model with suitably chosen parameter values:
$t_{life}=2.0\times 10^{7}$ yrs, $\xi=0.10 \ M_{\odot}/L_{\odot,B}$,
and $M_{min}=2.8\times 10^{10} \ M_{\odot}$. In Figure 5, we plot the
resulting number of sources invoked as a function of redshift.
In this model, the choice for the source lifetime simply reflects a
reasonable guess based on the quasar light curve derived by Haiman \&
Loeb (1998) for a similar cosmology and LF. Our value for the minimum
mass is set to the minimum group mass below which our simulation can
no longer resolve collapsed objects. We infer this mass limit from the
clear drop off in the low end of the mass function occurring at
roughly 64 gas particles or $2.8\times 10^{10} \ M_{\odot}$
(see, also, Katz et al. 1999; Weinberg et al. 2000).  With the
minimum and maximum mass of the groups identified, a mass-to-light
ratio of $\xi=0.10 \ M_{\odot}/L_{\odot,B}$ is then chosen to
reproduce roughly the same range of observed quasar luminosities from
the survey conducted by Warren, Hewett \& Osmer (1994). Consequently,
our ability to properly match the high end of the luminosity function
with our most massive groups ensures that our volume is large enough
to properly sample the range of source masses. It is important to
point out that although we have matched the observed range in
luminosities, there are undoubtably many more sources that are too
faint to have been observed that we are excluding from our
analysis. In fact, the results of Cheng et al. (1985) show that the LF
of Seyfert galaxies (which are well correlated with that of optically
selected quasars at $M_B=-23$) does not show clear evidence of
leveling off down to $M_B\simeq -18.5$ or $L_{min}\simeq 6.44\times
10^{9} \ L_{B,\odot}$, a factor of roughly $15$ below our choice for
$L_{min}$. Though these faint sources have a sizable contribution to
the total emissivity at low redshifts, the majority of ionizing
photons at high redshifts arise from the brightest (observed)
sources. In particular, at $z=3.3$ the luminosity function used in
this paper yields only a $12\%$ deficit in the total number of ionizing
photons from quasars when considering sources with luminosities
greater than $15$ times the minimum luminosity implied by Cheng et
al. (1985). Thus for the sake of being able to match the observed
range in luminosities with a constant mass-to-light ratio and avoid
accounting for the plethora of dim sources, we use a minimum
luminosity of $L_{min}=M_{min}/\xi=2.8\times10^{11} \ L_{B,\odot}$.
We do not expect the small fractional difference in the ionizing
emissivity predicted using this inflated $L_{min}$ to significantly
alter the morphological evolution of the reionization process.

\subsection{Evolution Code}

In Table 1, we list some of the variables necessary to properly track
the ionization state in a cell whose contents change dynamically in
time. In this section we describe in detail the computational steps
involved in determining the evolution of these parameters during the
reionization calculation.

We begin by first retrieving the gas density, $\rho$, the clumping
factor $C_f$, the recombination rates $\alpha_{\text{\tiny{He {\tiny
{II}}}}}^A$ and $\alpha_{\text{\tiny{He {\tiny {II}}}}}^B$ , and the
collisional ionization rate $\Gamma^{{\tiny coll}}_{\text{\tiny{He
{\tiny {II}}}}}$, for the earliest redshift $z_i$ when sources are due
to switch on. (Note since $\alpha_{\text{\tiny{He {\tiny {II}}}}}^A$
and $C_f$ or $\alpha_{\text{\tiny{He {\tiny {II}}}}}^B$ and $C_f$ will
always appear as a product in all our expressions, we conserve memory
by combining them into the variables $\alpha_{\text{\tiny{He{\tiny
{II}}}}}^AC_f$ and $\alpha_{\text{\tiny{He{\tiny {II}}}}}^BC_f$
respectively.  Similarly, the collisional ionization rate and the gas
clumping factor is combined into the variable $\Gamma^{{\tiny
coll}}_{\text{\tiny{He {\tiny {II}}}}}C_f$.)  Having already
calculated $\rho$, $\alpha_{\text{\tiny{He{\tiny {II}}}}}^AC_f$,
$\alpha_{\text{\tiny{He{\tiny {II}}}}}^BC_f$, and $\Gamma^{{\tiny
coll}}_{\text{\tiny{He {\tiny {II}}}}}C_f$ for every hydro output, we
use information from two adjacent output grids at redshifts $z_1$ and
$z_2$, where $z_1>z_i>z_2$, to linearly interpolate the variables for
our grid at $z_i$. In particular, for each cell we calculate
derivatives for $\rho$, $\alpha_{\text{\tiny{He{\tiny {II}}}}}^AC_f$,
$\alpha_{\text{\tiny{He{\tiny {II}}}}}^BC_f$, and $\Gamma^{{\tiny
coll}}_{\text{\tiny{He {\tiny {II}}}}}C_f$ between redshifts $z_1$ and
$z_2$ and store the results in $\partial \rho/\partial z$, $\partial
(\alpha_{\text{\tiny{He{\tiny {II}}}}}^AC_f)/\partial z$, $\partial
(\alpha_{\text{\tiny{He{\tiny {II}}}}}^BC_f)/\partial z$, and
$\partial (\Gamma^{{\tiny coll}}_{\text{\tiny{He {\tiny
{II}}}}}C_f)/\partial z$ respectively. These derivatives are then used
to linearly interpolate the values of the variables at the
intermediate redshifts.

Having retrieved $\rho$ we initialize the variable
$n_{\text{\tiny{He}}_{\text{tot}}}$ which stores the comoving number
density of helium nuclei,
\begin{equation}
n_{\text{\tiny{He}}_{\text{tot}}}=\frac{\rho \
Y}{m_{\text{\tiny{He}}}},
\end{equation}
where $Y$ is the helium abundance by mass (we adopt $Y=0.24$ and
$X=0.76$ for hydrogen from the hydro simulations) and
$m_{\text{\tiny{He}}}$ is the mass of the helium atom. To initialize
the number density of electrons $n_e$, we will first need to assume an
initial ionization state of the gas at the start of our
simulation. Applications of the Gunn-Peterson constraint have shown
that the hydrogen component of the intergalactic gas was in a highly
ionized state by $z\sim 5$ (Schneider, Schmidt, \& Gunn 1991) or even
$z\sim 6$ (Fan et al. 2000) and that almost all the helium in the
universe was singly ionized by about the same time. Due to the sharp
decline in the number density of bright quasars for $z>5$, it appears
as if either low-luminosity quasars or first generation stellar
sources were responsible for the ionization of these components of the
intergalactic gas. In any case, since the first sources of He
{\small{II}} ionizing flux in our simulation switch on around $z\sim
5$, we simply initialize the ionization fraction for H {\small{II}}
and He {\small{II}} to unity. We are now able to compute the initial
comoving electron density,
\begin{equation}
n_e=\rho\
\biggl(\frac{X}{m_{\text{\tiny{H}}}}+\frac{Y}{m_{\text{\tiny{He}}}}\biggl)\
+\ \chi_{\text{\tiny{He II}}}n_{\text{\tiny{He}}_{\text{tot}}},
\end{equation}
where the first two terms on the right hand side represent the free
electrons contributed from H {\small{II}} and He {\small{II}}
respectively, and where the last term represents electrons released
from He {\small{II}} ionizations. Since we have initialized
$\chi_{\text{\tiny{He II}}}=0$, our initial value for the electron
density will come only from the first two terms.

Once variables are initialized we can proceed with the ionization
calculation. The computation will progress in uniform time increments
corresponding to source redshift bins $z_i$. At each $z_i$, a
predetermined number of sources will turn on throughout the box with
$\chi_{\text{\tiny{He II}}}$ and $n_e$ being updated for each
source. For each $z_i$, precalculated derivatives in each cell are
used to interpolate new densities and rates, with derivatives being
updated near every hydro output redshift.

To calculate the ionization fractions we require an expression for the
cross section of the species which is implicit in the photoionization
rate.  Using the smoothed interpolation formula for the
photoionization cross section of He {\small II} given in Osterbrock
(1989; eq. [2.31]) and assuming that $J(\nu)\propto \nu^{-1.8}$ beyond
the ionization threshold frequency $\nu_I$, we arrive at a mean cross
section of $\bar{\sigma}_{\text{\tiny{He II}}}=0.395 \
\sigma_{\text{\tiny{He II}}}(\nu_I)$.  The numerical value of
$\sigma_{\text{\tiny{He II}}}(\nu_I)$ is taken from Osterbrock (1989;
eq. [2.4]).

\subsection{Results}

In Figure 6, we show a series of snapshots of a slice through our
simulation volume demonstrating how the reionization process develops
for our particular model. In each snapshot, ionized regions with
$\chi_{\text{\tiny{He II}}}>0.90$ are outlined by the solid contours,
with the underlying greyscale image indicating the density field. One
can notice how the fronts quickly expand first into the voids, then
more slowly into the denser filaments. Conversely, denser regions are
the first to recombine after they are mostly ionized due to the fact
that they have large clumping factors. Snapshots such as these can
allow for qualitative comparisons between a number of models which can
then be linked to quantitative differences.

To examine the manner in which each model ionizes the IGM, we track
global parameters describing the ionization fraction of our box at
each redshift. Figure 7 shows how the ionized mass fraction (solid
line) and the ionized volume fraction (dashed line) evolve with
redshift.  The entire volume can become fully ionized by
$z\sim3.3$, consistent with observations (Jakobsen et al. 1994;
Davidsen, Kriss, \& Zheng 1996; Hogan, Anderson, \& Rugers 1997;
Reimers et al. 1997; Heap et al. 2000; Smette et al. 2000). It is
evident that whereas the volume can easily achieve full
ionization, the mass component cannot. This is a consequence of
extremely overdense cells with large clumping factors quickly
recombining after they have been ionized.  Due to their high
overdensities (factors of $10^3$ or more) it takes relatively few such
cells to substantially lower the ionized mass fraction. Furthermore,
one expects the volume fraction to quickly exceed the mass fraction as
I-fronts seep into the more voluminous underdense regions, which is
also evident from the plot.

\section{CONCLUDING REMARKS}

In light of recent observations of the spectra of high-redshift
quasars, a detailed understanding of the thermal history of the IGM
and reionization is highly desirable. In this paper we have described
a method that is able to simulate the inhomogeneous reionization of a
cosmological volume from a set of point sources and a diffuse
background component. This new method conceptually differs
significantly from previous approaches in that it does not evolve the
radiation on the opacity timescale but rather on source lifetimes
(ANM99; Ciardi et al. 2000; Razoumov and Scott 1999; Gnedin 2000;
Kessel-Deynet and Burkert 2000; Norman, Paschos \& Abel 1998).  Our
method is specifically designed with short-lived sources in mind where
one expects a negligible amount of cosmological evolution during a
source lifetime.  However, it is feasible to extend the formalism
described in this paper to long-lived sources as well by solving the
radiative calculations in a series of sub-steps, each of which is
comparable to the lifetimes of short-lived sources where the
assumption of negligible cosmological evolution is still valid.

As a first application, we have used our algorithm to conduct a
statistical study of the reionization of He {\small{II}} from
short-lived quasars. Our main conclusions from this application are:
(1) quasars with the spectral form given in Madau, Haardt, \& Rees
(1999) are able to ionize the volume by a redshift $\simeq 3.3$ for a
reasonable choice of source characteristics, and (2) that ionization
fronts preferentially expand into the least dense but more voluminous
regions of the IGM, resulting in the ionized volume fraction being
consistently larger than the ionized mass fraction. This feature seems
to be less pronounced when more sources are involved since their
presence in overdense regions counteracts the effect. These findings
are in accord with the expectations from the model for reionization of
Miralda--Escud{\'e}, Haehnelt, and Rees (2000).

Our analysis in this paper reflects a first step in applying our
algorithm to address a large assortment of questions regarding
radiation processes in the universe. In particular, the method we have
presented here can be supplemented to include the effect of heating
from photoionizations to more accurately study the thermal history of
the IGM. In addition, one can extract simulated Ly$\alpha$ spectra
from these volumes, which can then be used as a comparison tool with
observational results.  We can also envision integrating our algorithm
within a hydrodynamical cosmological simulation to study the effects
of ionizations on the dynamics of matter on various scales.  A more
systematic study of radiative transfer effects for He reionization by
short-lived quasars will be presented in a forthcoming paper.

\begin{acknowledgments}

We thank Volker Springel for making the outputs from the cosmological
simulation available to us in a convenient form as well as providing
useful algorithms for handling the large datasets. We thank Abraham
Loeb for stimulating and informative discussions regarding the
reionization of the universe. We also thank Zolt$\acute{\text{a}}$n
Haiman for providing us with a useful program for computing luminosity
functions as well as informative commentary.  This work was supported
in part by NSF grants ACI96-19019, AST-9803137, and PHY 9507695.

\end{acknowledgments}

\clearpage

\begin{table}[htb]
\begin{center}
\begin{tabular} {ll}
\multicolumn{2}{c}{\textbf{TABLE 1 }} \\
\multicolumn{2}{c}{\small{Variable List for Tracking Cell Evolution}}
\\ \hline \hline Variable & Description \\ \hline 
$\rho$.......................... & comoving gas density \\

$n_{\text{\tiny{He}}_{\text{\tiny{tot}}}}$ .................... &
comoving number density of helium nuclei \\ 

$n_e$........................ & comoving electron number density \\

$\alpha_{\text{\tiny{He {\tiny {II}}}}}^AC_f$ ................ & gas
clumping factor $\times$ Case A recombination rate \\ 

$\alpha_{\text{\tiny{He {\tiny {II}}}}}^BC_f$ ................ & gas
clumping factor $\times$ Case B recombination rate \\ 

$\Gamma^{{\tiny coll}}_{\text{\tiny{He {\tiny {II}}}}}C_f$ .................... & gas
clumping factor $\times$ collisional ionization rate \\

$\chi_{\text{\tiny{He III}}}$ ................... & He \small{III}
ionization fraction ($n_{\text{\tiny{He III}}}/n_{\text{\tiny{He}}_{\text{\tiny{tot}}}}$) \\ 
$\partial\rho/\partial z$ .................. & rate of change of comoving gas
density with redshift \\ 

$\partial {\small{(\alpha_{\text{\tiny{He
{\tiny {II}}}}}^AC_f)}}/\partial z$ ...... & rate of change of product
$\alpha_{\text{\tiny{He {\tiny {II}}}}}^AC_f$ with redshift \\ 

$\partial {\small{(\alpha_{\text{\tiny{He
{\tiny {II}}}}}^BC_f)}}/\partial z$ ...... & rate of change of product
$\alpha_{\text{\tiny{He {\tiny {II}}}}}^BC_f$ with redshift \\ 

$\partial(\Gamma^{{\tiny coll}}_{\text{\tiny{He {\tiny {II}}}}}C_f)/\partial z$
............. & rate of change of product$\Gamma^{{\tiny coll}}_{\text{\tiny{He {\tiny {II}}}}}C_f$ with redshift\\

\hline
\end{tabular}
\end{center}
\end{table}

\clearpage

\begin{figure}[htb]
\figurenum{1}
\setlength{\unitlength}{1in}
\begin{picture}(6,7.0)
\put(-1.0,-2.0){\includegraphics{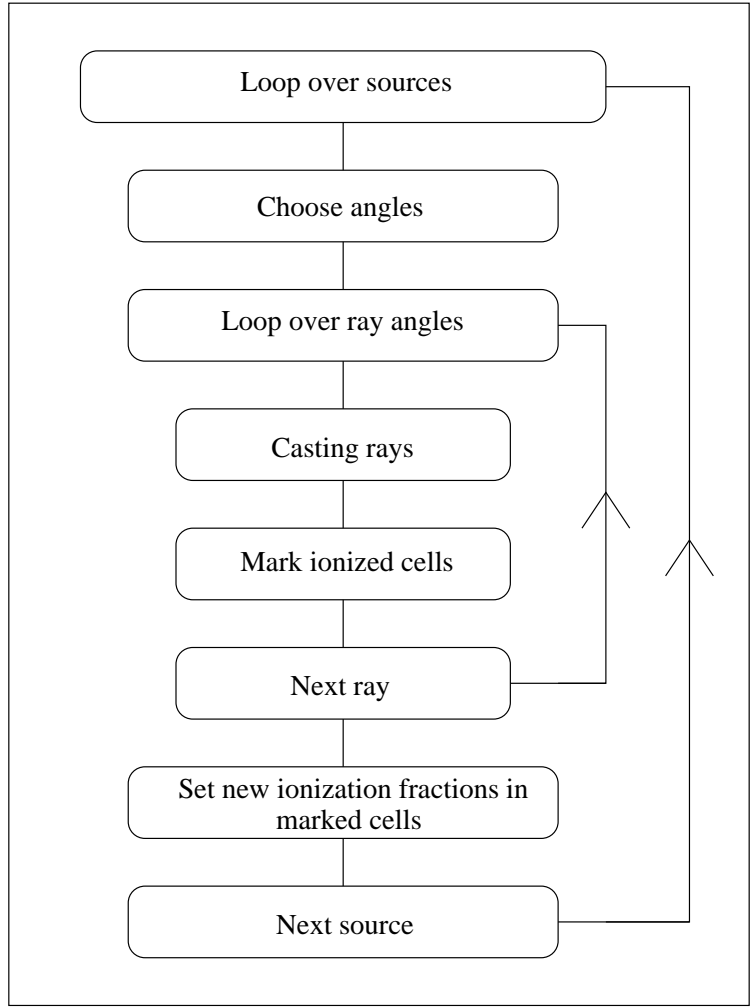}}
\end{picture}
\caption{Block diagram for ionization algorithm used for sources.}
\end{figure}

\clearpage

\begin{figure}[htb]
\figurenum{2}
\setlength{\unitlength}{1in}
\begin{picture}(6,7.0)
\put(0.0,-1.0){\includegraphics{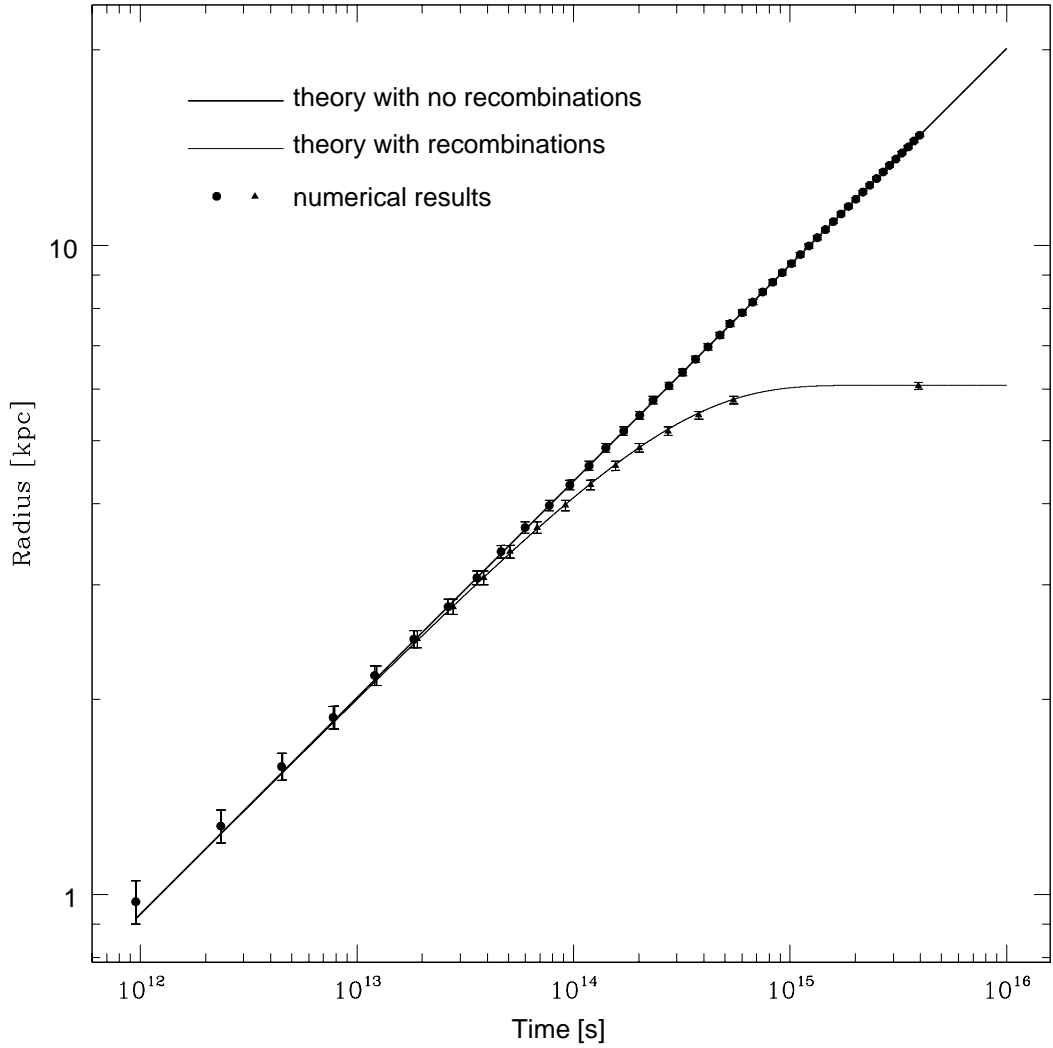}}
\end{picture}

\caption{Comparison of analytic radius of an I-front with the
numerical solution on a $100^3$ uniform grid in case where
recombinations are included (triangles) and excluded (circles).  The
error bars represent the maximum deviation found for the spherical
ionization front and are always less than one grid cell.}

\end{figure}

\clearpage

\begin{figure}[htb]
\figurenum{3}
\setlength{\unitlength}{1in}
\begin{picture}(6,7.0)
\put(-1.0,-2.0){\includegraphics{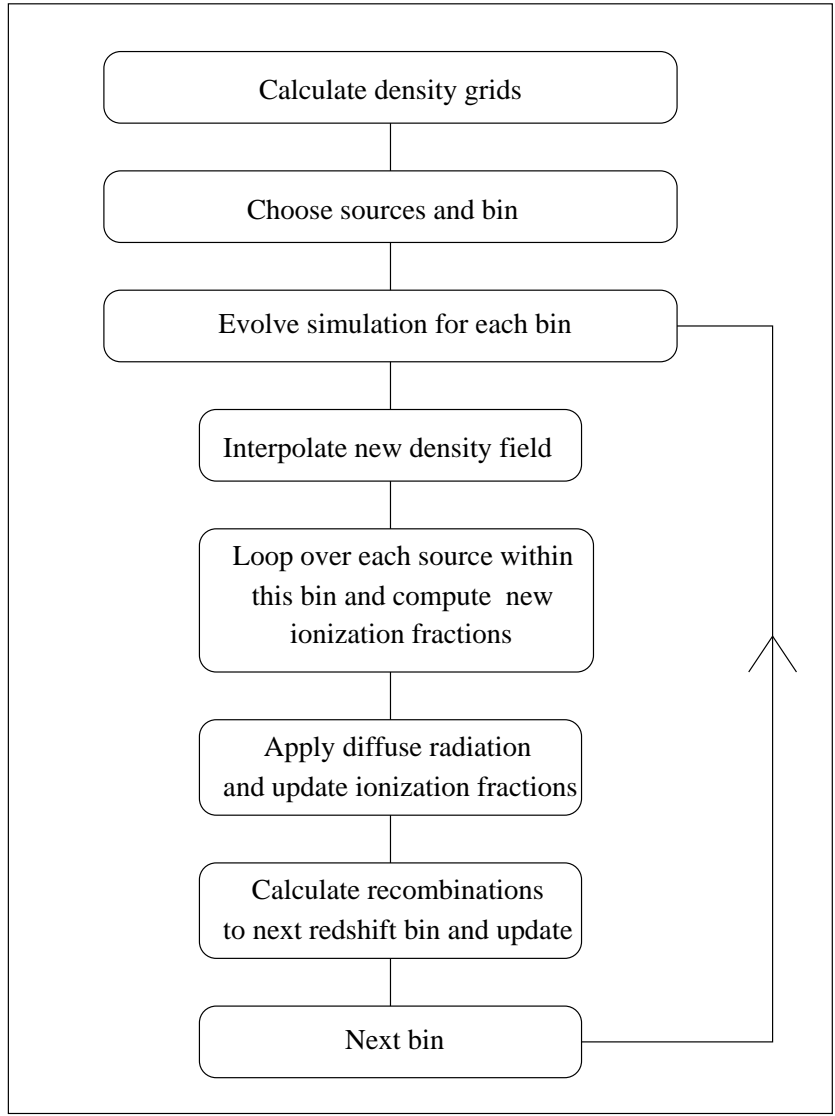}}
\end{picture}
\caption{Block diagram for the reionization simulation.}
\end{figure}

\clearpage

\begin{figure}[htb]
\figurenum{4}
\setlength{\unitlength}{1in}
\begin{picture}(6,7.0)
\put(-1.0,-2.0){\includegraphics{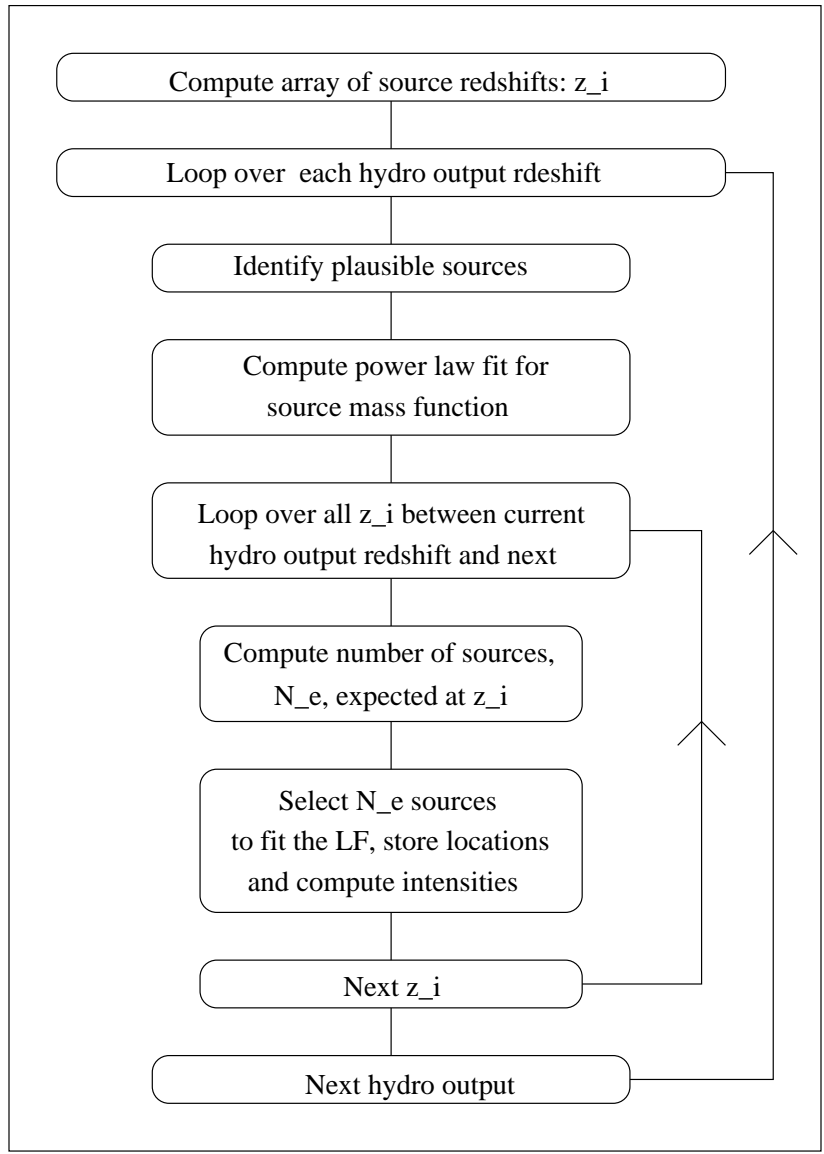}}
\end{picture}
\caption{Block diagram for source selection analysis.}
\end{figure}
\clearpage

\begin{figure}[htb]
\figurenum{5}
\setlength{\unitlength}{1in}
\begin{picture}(6,6.5)
\put(-0.0,-0.5){\includegraphics{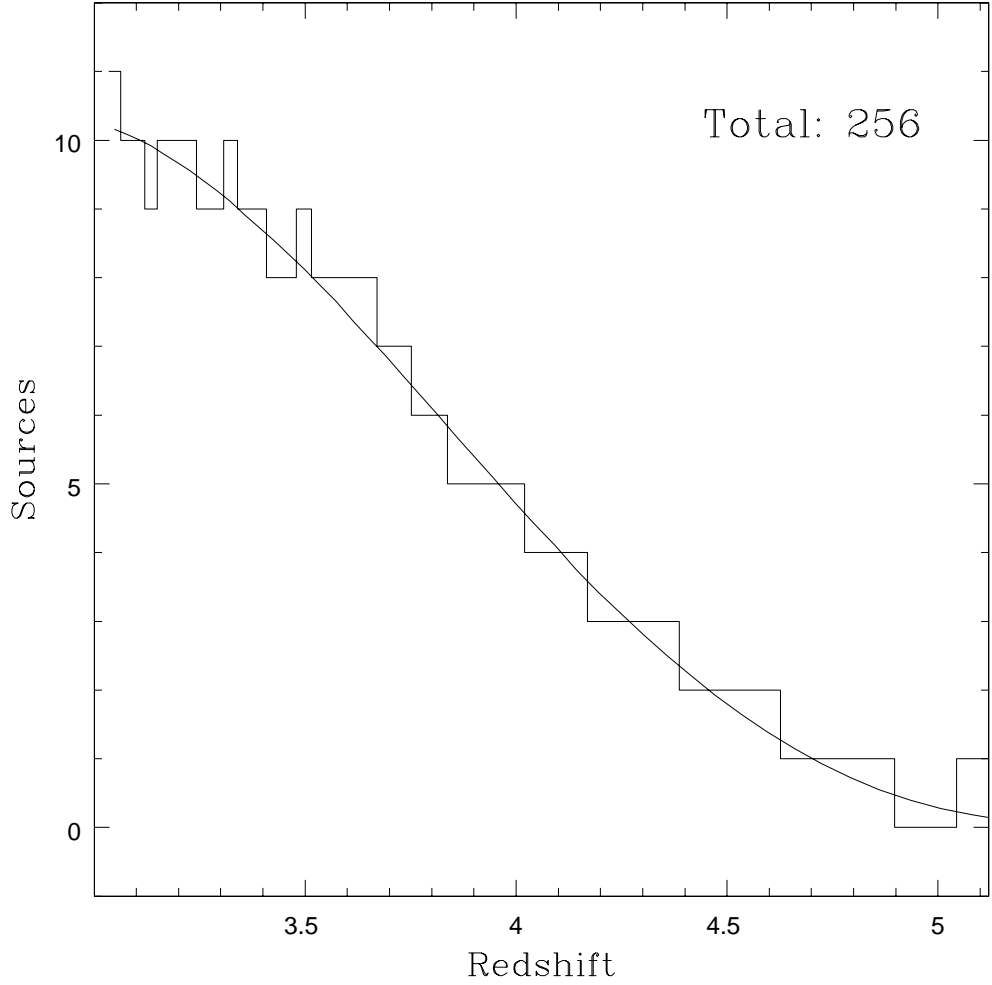}}
\end{picture}

\caption{Active source count versus redshift. The solid curve
represents the number expected according to the LF (see eq. 32). All
binsizes correspond to a source lifetime of $3.0 \times 10^{7}$ yrs. A
total of 46 sources were invoked between $3.0< z <5.7$.}

\end{figure}

\clearpage

\begin{figure}[htb]
\figurenum{6}
\setlength{\unitlength}{1in}
\begin{picture}(6,6.3)
\put(-0.09,-0.9){\includegraphics{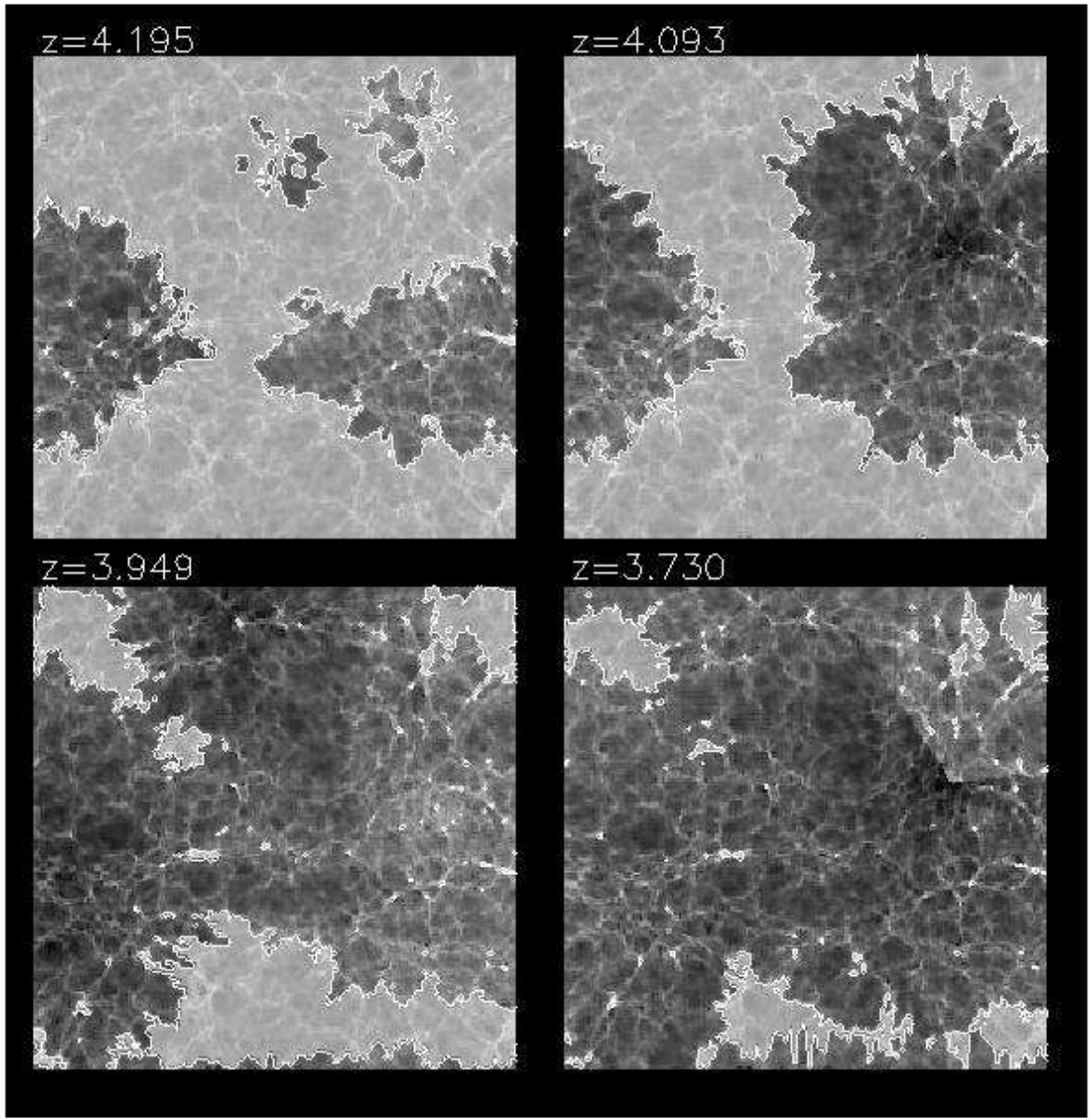}}
\end{picture}

\caption{Snapshots of a slice through our simulation volume at
$z=4.195, 4.093, 3.949$, and $3.739$ ({\it from top left to bottom
right}). Solid contours indicate regions with $\chi_{\text{\tiny{He
II}}}>0.90$ with the underlying greyscale image representing the 
logarithmic
density of He {\small II} ions . Note that as sources turn on, the 
ionization zones preferentially expand into regions of low density.}

\end{figure}

\clearpage

\begin{figure}[htb]
\figurenum{7}
\setlength{\unitlength}{1in}
\begin{picture}(6,6.5)
\put(.2,-0.8){\includegraphics{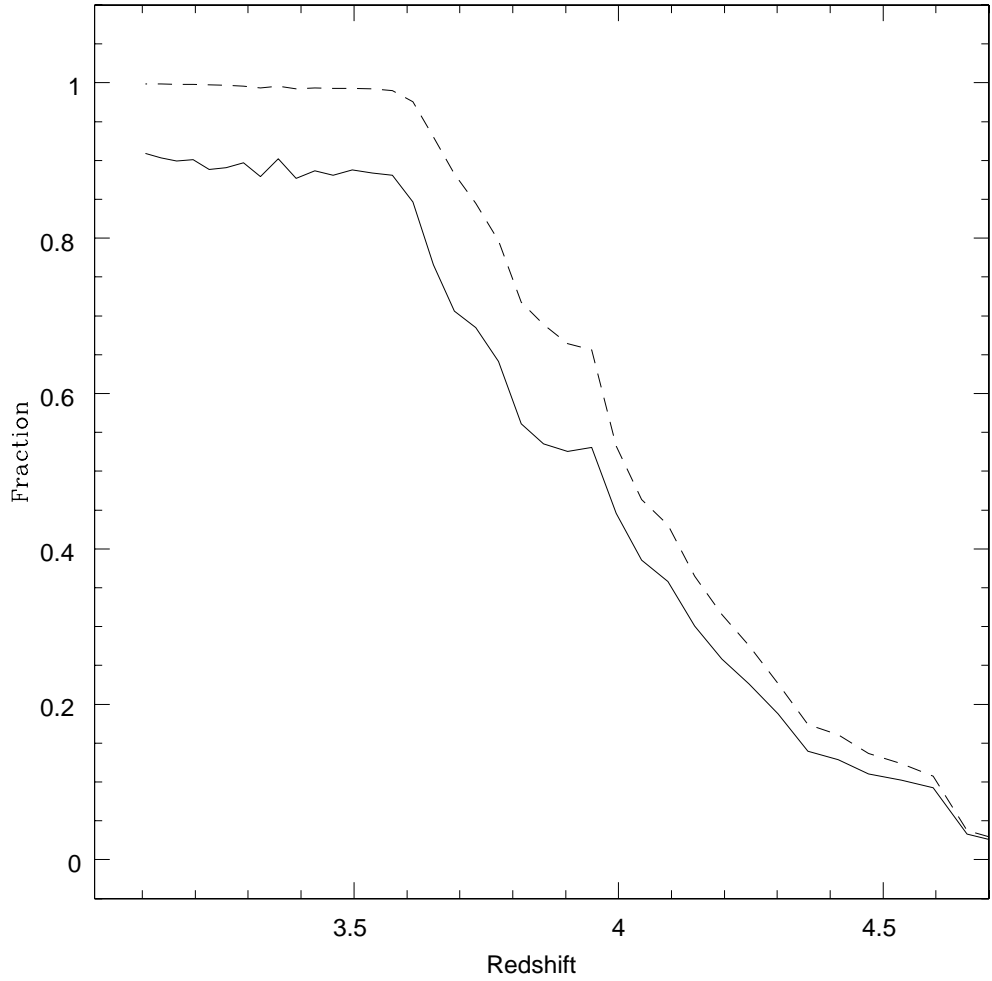}}
\end{picture}

\caption{Ionized mass fraction (solid line) and ionized volume
fraction (dashed line) as a function of redshift. In this model the
volume becomes fully ionized by a redshift of $z\simeq 3.3$ with the
volume fraction being consistently larger than the mass fraction (see
text for discussion).}

\end{figure}

\end{document}